\documentclass[%
reprint,
normalem,
aps,
prb,
]{revtex4-2}

\usepackage{amsmath,amssymb}
\usepackage{graphicx}

\usepackage[caption=false]{subfig}

\usepackage[pdftex, hidelinks,
pdfauthor={Johannes Mellak},
pdftitle={no title},
pdfproducer={Latex with hyperref, or other system},
pdfcreator={pdflatex, or other tool}]{hyperref}

\usepackage{bm}  
\newcommand{\ve}[1]{{\bm{#1}}} 
\newcommand{\rh}{\hat{\rho}}
\newcommand{\si}{{\bm{\sigma}}}
\newcommand{\Li}{{\mathcal{L}}}
\newcommand{\vtheta}{{\bm{\theta}}}

\usepackage{bbold}
\usepackage{color}

\usepackage{cancel}
\usepackage{ulem}
\usepackage[dvipsnames]{xcolor}

\begin{document}

\title{Deep neural networks as variational solutions for 
correlated open quantum  systems
} 

\author{Johannes Mellak}
\email{mellak@tugraz.at}
\author{Enrico Arrigoni}
\author{Wolfgang von der Linden}
\affiliation{Institute of Theoretical and Computational Physics, Graz University of Technology, Petersgasse 16/II
	A-8010 Graz, Austria}

\date{\today}

\begin{abstract}

In this work we apply deep neural networks to find the non-equilibrium steady state solution to 
correlated open quantum many-body systems. 
Motivated by the ongoing search to find more powerful representations of (mixed) quantum states, we design a simple prototypical 
convolutional neural network and show that parametrizing the density matrix directly with more powerful models can yield better variational ansatz functions and improve upon results reached by neural density operator based on the restricted Boltzmann machine. 
Hereby we give up the explicit restriction to positive semi-definite density matrices. 
However, this is fulfilled again to good approximation by optimizing the parameters. 
The great advantage of this approach is that it opens up the possibility of exploring more complex network architectures that can be tailored to specific physical properties. 
We show how translation invariance can be enforced effortlessly and reach better results with fewer parameters. We present results for the dissipative one-dimensional transverse-field Ising model and a two-dimensional dissipative Heisenberg model compared to exact values. 

\end{abstract}

\maketitle

\section{Introduction}


Finding solutions to strongly correlated quantum many-body systems, where the Hilbert space comprising all possible configurations grows exponentially with system size, 
relies on approximations and numerical simulation. 
In recent years neural networks encoding quantum many-body states, termed neural network quantum states (NQS), 
have emerged as a promising tool  for a variational treatment of quantum many-body problems \cite{carleo_solving_2017}. 
%
%
Using 
gradient-based optimization, the groundstate or time-evolved 
quantum wavefunctions as solutions to the many-body Schrödinger equation 
can be efficiently approximated \cite{carleo_solving_2017, hibat-allah_recurrent_2020, choo_study_2019, choo_fermionic_2019, schmitt_quantum_2020, hermann_deep-neural-network_2020, pfau_ab_2020, gerard_gold-standard_2022, carrasquilla_how_2021, roth_high-accuracy_2023},  
exploiting the expressive powers and the universal approximation properties of state of the art machine learning techniques that also often face very high dimensional problems. 
%
%
%
%
These ansatz functions often require less parameters to express the exponential complexity of 
correlations 
and bear the prospect of an improved scaling behaviour to larger and higher dimensional systems. 
%

Open quantum systems, where the system is connected to an environment or \emph{bath} that induces dissipative processes described by a quantum master equation, have become of great interest  
in part due to the advent of  
noisy quantum devices such as 
quantum computers. %
Neural network density operators (NDO), parameterizing the mixed density operator describing such systems, 
have recently been shown to be a capable numerical tool to compute the dynamics of open quantum systems 
 \cite{torlai_latent_2018, hartmann_neural-network_2019, nagy_variational_2019, vicentini_variational_2019, yoshioka_constructing_2019}. %
%
%
So far there are mainly two approaches. 
One is to analytically trace out additional bath degrees of freedom in a purified restricted Boltzmann machine (RBM) \cite{torlai_latent_2018}, yielding an ansatz function that always fulfils all the properties of a physical density matrix, as used in \cite{vicentini_variational_2019, hartmann_neural-network_2019, kaestle_sampling_2021, mellak_quantum_2023}. %
An alternative approach is to describe the system using a probabilistic formulation via positive operator-valued measurements (POVM) \cite{schmale_efficient_2022, PhysRevLett.128.090501}. 
This method works with arbitrary network ansatz functions and 
%
was shown to improve upon results reached by purified RBM for some cases \cite{PhysRevLett.128.090501, zhao_empirical_2023}. %



We will focus on an improvement of the first method 
and address the constraints which limits the choice of parametrization. 
For a matrix to represent a physical density operator, it must be positive semidefinite and Hermitian,
which the purified ansatz using RBM fulfils by construction. 
If the positivity condition is not enforced however, 
this opens up new possibilities to use more powerful representations. 
More modern deep neural networks were shown to outperform shallow RBM and fully connected networks for closed quantum systems, where the complex groundstate wavefunction is targeted instead of a mixed density matrix, 
f.e. \cite{gao_efficient_2017, levine_quantum_2019}. 

In the ongoing search for better variational functions to approximate mixed density matrices, 
in this work we apply prototypical deep convolutional neural networks (CNN), which are part of most modern deep learning models, 
to open quantum systems, by parametrizing the density matrix directly. 
Due to the parameter sharing properties of CNN, translation symmetry can be enforced easily. %
This also leads to a system size independent parametrization, enabling transfer learning to larger systems. 
We find considerably improved results compared to NDO based on RBM, using much fewer parameters. 

In this paper we describe the formulation of finding the steady state solution to the Lindblad master equation as an optimization problem. Then we introduce a neural network architecture based on convolutional neural networks to encode the density matrix of translational invariant open quantum systems and finally we present 
our 
results. 

\section{Optimizing for the nonequilibrium steady-state density operator}


The dynamics of an open quantum system with Hamiltonian $H$ coupled to a Markovian environment is described by the Lindblad master equation
\begin{equation}
	\frac{d\rh}{dt} = \Li \rh 
	= -i \left[H, \rh \right] 
	+ \sum_k \gamma_k 
	\big( L_k \rh L_k^\dagger - \frac{1}{2} \{ L_k^\dagger L_k, \rh \} \big)
	\label{eq:lindblad}
\end{equation}
with the 
\emph{jump} 
operators $L_k$ 
leading to non-unitary dissipation. 
The non-equilibrium steady-state (NESS) density matrix $\rh$, where
$d\rh / dt = 0$
which is reached in the long-time limit, 
can be obtained directly via a variational scheme by minimizing a norm 
of the time derivative in Eq.~(\ref{eq:lindblad}) \cite{weimer_variational_2015}. 

Neural networks as variational ansatz for the density operator $ \rh_{\vtheta} = \sum_{\si \si'} \rho_{\vtheta}(\si, \si') |\si\rangle\langle\si'|$ parametrized by the set of 
parameters $\vtheta$, 
with the complete many-body basis of spin-$1/2$ configurations $|\si\rangle$, 
have been used to find the NESS solution to different open quantum spin systems \cite{torlai_latent_2018, vicentini_variational_2019, nagy_variational_2019, hartmann_neural-network_2019,  yoshioka_constructing_2019}. 
We use the $L_2$ norm as the cost function to 
be minimized 
as was described in Ref.~\cite{vicentini_variational_2019} 
\begin{align}
	C(\vtheta) &= 
	\frac
	{\text{Tr} \left[\rh_{\vtheta}^\dagger \Li^\dagger \Li \rh_{\vtheta} \right]}
	{\text{Tr} \left[\rh_{\vtheta}^\dagger \rh_{\vtheta} \right]}
	= \frac
	{|| \Li \rh_{\vtheta}||^2_2}
	{||\rh_{\vtheta}||^2_2}\\
	&= \sum_{\si\si'} p_\vtheta (\si,\si') 
	\left| \sum_{\tilde{\si}\tilde{\si}'}	\Li_{\si\si'\tilde{\si}\tilde{\si}' }
	\frac{\rho_\theta(\tilde{\si}, \tilde{\si}')}
	{\rho_\theta(\si, \si')}\right|^2 
	\; .
	\label{eq:sum:HS} 
\end{align}
This function and its derivative with respect to the parameters $\vtheta$ 
can then be evaluated as the statistical expectation value over the probability distribution
\begin{align}
	p_\vtheta (\si,\si')  = \frac{|\rho_\theta(\si, \si')|^2}{\sum_{\bar{\si}\bar{\si}'}| \rho_\theta(\bar{\si}, \bar{\si}')|^2}
\label{eq:prob_sig_sig'}
\end{align}
using Monte-Carlo samples. 
This avoids the first sum over the entire Hilbert space, while the inner sum in Eq.~\eqref{eq:sum:HS},
 which adds up the sparse Lindblad matrix elements $\Li_{\si\si'\tilde{\si}\tilde{\si}' }$, 
 can usually be carried out exactly. 
The parameters are then iteratively updated to find the steady state as the solution to the optimization problem 
\begin{equation}
\rho_{SS} = \underset{\rho_{\vtheta}}{\text{argmin}} \, C(\vtheta) \; .
\end{equation}

To update the parameters, the stochastic reconfiguration (SR) method \cite{sorella_weak_2007} is often used for optimizing neural quantum states \cite{carleo_solving_2017}, 
where a system of equations is solved in each iteration to adapt the metric to the current cost surface. 
However, we find improved convergence and often better results using a backtracking Nesterov accelerated gradient descent optimization scheme as described in Ref.~\cite{mellak_quantum_2023} (NAGD+),  especially when optimizing deeper neural networks. 
With automatic differentiation, 
the gradients of Eq.~\eqref{eq:sum:HS} are evaluated using the same Monte-Carlo samples. 

Once the optimization is converged, for a given set of parameters the expectation values of physical observables $\hat{O}$ are computed as an expectation value 
	\begin{equation}
		\label{eq:observable}
		\langle \hat{O} \rangle = \text{Tr} \; \{\hat{O} \rho \}
		= \sum_\si p(\si) \sum_{\si'} \frac{\langle \si | \hat{O} | \si'\rangle \rho(\si, \si')}{\rho(\si, \si)} 
	\end{equation}
over Monte-Carlo samples from the probability distribution $p(\si) = \rho(\si, \si)$. 
This way, the summands are independent of the normalization of $\rho$. 
Again the inner sum can typically be carried out exactly. 
A large part of computational effort is taken up by the Monte-Carlo sampling, which, however, can be highly parallelized and in some cases accelerated by implicitly enforcing symmetries during sampling \cite{mellak_quantum_2023}.

\section{Deep Neural network density operator ansatz}
\label{sec:network}



\subsection{Neural-network density operator based on purification}

The neural-network density operator (NDO) based on 
a purified 
RBM is defined by analytically tracing out additional \emph{ancillary} nodes $\ve{a}$ in an extended system described by a parametrized wavefunction $\psi(\si, \ve{a})$ \cite{torlai_latent_2018}. 
The reduced density matrix for the physical spin configurations $\si$, $\si'$ 
is then obtained by marginalizing over these bath degrees of freedom $\ve{a}$ 
	\begin{equation}
		\label{eq:rho_ansatz}
		\rho(\si, \si') = \sum_{\ve{a}} \psi^*(\si, \ve{a}) \psi(\si', \ve{a}) 
		\; .
\end{equation}
This can be done as long as the dependence on $\ve{a}$ in $\psi$ can be factored out, as is the case for the RBM wavefunction \cite{carleo_solving_2017} ansatz $\psi(\si, \ve{a}) = \exp{[-E(\si, \ve{a})]}$ 
adopted in Ref.~\cite{hartmann_neural-network_2019, nagy_variational_2019, vicentini_variational_2019} 
where $E$ is an Ising type interaction energy and thus a linear function of $\ve{a}$. 
But this particular design does not represent the most general density matrix, where the ancillary bath degrees of freedom would be 
	spins in the visible layer that are traced out, instead of another set of hidden nodes. 
When trying to extend the purification approach to deeper networks, the ancillary nodes cannot be traced out analytically any more, requiring a computationally expensive sampling of hidden layers \cite{nomura_purifying_2021}. 

On the other hand, 
these purified RBM NDO which represent a reduced density matrix in an extended system have the advantage of being positive-semidefinite and Hermitian by design. 
However, a variational ansatz encoding the density operator does not necessarily need to enforce these properties exactly, 
as was previously shown \cite{cui_variational_2015, yoshioka_constructing_2019}. 
Once the optimization problem yielding the steadystate solution is solved, the resulting density matrix should have these properties within some approximation error. 
Letting the optimization deal with this 
enables us to use new classes of potentially more expressive networks, such as CNN, as variational tools for open quantum systems. 

\subsection{Convolutional neural networks}
There have been multiple results showing that \emph{deep} network architectures, which have more than one  hidden layer of non-linear transformations, offer considerable advantages in expressing highly entangled quantum wavefunctions compared to shallow networks like RBM \cite{gao_efficient_2017, levine_quantum_2019}. 
Especially CNN have been applied very successfully to closed quantum many-body systems 
and problems in continuous space 
\cite{saito_machine_2018, choo_study_2019, pescia_neural-network_2022, roth_high-accuracy_2023}. %

Convolutional neural networks
are used in most modern neural network architectures, for example in image recognition \cite{lecun_deep_2015}. 
They work by applying convolution filters, which constitute a part of the variational parameters, to the input data, 
followed by a non-linear activation function and possibly some \emph{pooling} or averaging to decrease the feature size \cite{lecun_handwritten_1989}. 
The output of the $n$-th 
convolutional layer is computed as 
\begin{equation}
	F^{(n)}_{i, j, k} = f \left(\sum_{x=1}^X \sum_{y=1}^Y \sum_{c=1}^C F^{(n-1)}_{i + x, j + y, c} K^{(n-1)}_{x, y, c, k} \right)
	\label{eq:convolution}
\end{equation}
with the $k$-th Kernel $K$ of size $(X, Y, C)$,  
a non-linear activation function $f$ and setting $F^{(0)}$ to be the input layer. 
Here the indices $x$ and $y$ run over the spatial dimensions, 
whereas $c$ indexes the channels, i.e. how many kernels there were in the previous layer. 
In the case of image recognition for example, 
the channels dimension in the first layer is used to encode the colour channel in RGB input images.
In our case of spin configuration, the channels describe
 the left and right Hilbert spaces of the density matrix, as discussed below. 
 
Essentially, for each layer a fixed size matrix of parameters is scanned over the input, and the output is an array of inner products of this kernel with the input at the respective positions. 
Such a convolutional layer  
produces so called feature maps $F$ as the output.  These 
indicate the locations where the convolution filters $K$ matched well with the corresponding part of the input \cite{lecun_handwritten_1989}. 
From this it is apparent that convolutional kernels extract only 
local information or short range 
correlations between the input nodes for that layer.
However, successively applying multiple such layers, the field of view of the output nodes is increased. 
A convolutional layer can also be understood as a fully connected layer of neurons, where some parameters are shared between 
them, hence there are many more connections than parameters. 

\subsection{Convolutional neural-network density operator}
To represent a Hermitian density operator we start by 
parametrizing 
\begin{equation}
\rho_\vtheta(\si, \si') = A_\vtheta(\si, \si')^* + A_\vtheta(\si', \si)
\end{equation}
with the network output $A_{\vtheta}$ and a set of variational parameters $\vtheta$. 
It makes sense to consider the locality of the spins $\sigma_i$ and $\sigma'_i$ at site $i$ in the design of the network. 
This can be done by stacking $\si$ and $\si'$ instead of concatenating them, essentially introducing a new dimension. 
For a one-dimensional chain of $N$ spins, the input then becomes a 2D \emph{image} of size $N \times 2$
where the pairs $(\sigma_i, \sigma'_i)$ stay together. 
In common neural network software libraries 
the \emph{channels} dimension of the input layer can be used for this. 
Lattices with more than one dimension are equally easy to implement in this manner. 
We then apply two or more convolutional layers with fixed-size kernels to the input nodes. 
The kernel sizes together with the depth of the network determine how well long-range correlations can be represented. 
The resulting feature maps in each layer are transformed element-wise by 
the leaky variant of the rectified linear unit (ReLU), defined as 
\begin{equation}
	f(x) = \max[0, (1-\alpha) x] + \alpha x
\end{equation}
with $\alpha=0.01$. 
To obtain a complex density matrix amplitude $A$, 
the final feature maps 
are taken as the input to a fully connected layer with two output neurons $F_{(0,1)}$ 
representing the real and imaginary parts of $A = F_{\text{0}} + i F_{\text{1}}$. 
In this way, 
 all computations inside the model can be done using real-valued parameters. 
In terms of Eq.~\eqref{eq:convolution} this step can be understood as applying two kernels of the same size as the previous layer's output. 

\begin{figure}[h!]
	\centering
	\includegraphics[clip, trim=0cm 0cm 0cm 0cm,
	width=1\linewidth]
	{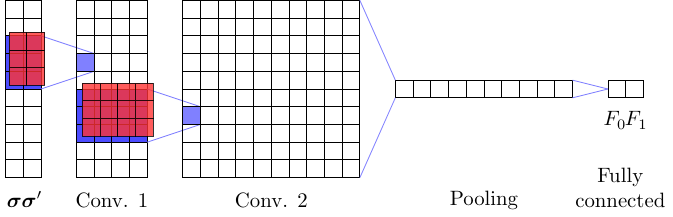} 
	\caption{Architecture of a simple CNN network ansatz for translation invariant density matrices of a one dimensional system: 
			Two convolutional layers with $4$ and $10$ kernels of shape 
			$(X,Y,C)=(3,1,2)$ 
			and $(3,1,4)$ respectively, indicated in red. 
			The columns represent the $\si$ and $\si'$ input vectors in the first layer and the feature maps in the consequent layers. 
			For translation invariant models, a pooling layer averages over the spatial dimensions. Finally a fully connected layer with two output nodes $F_0$ and $F_1$ map to the real and imaginary part of the density matrix element. For a two dimensional lattice, we simply introduce an additional dimension in the input and all convolutional layers. 
		}
	\label{fig:architecture}
	\hfill
\end{figure}

The variational parameters consist of the convolution kernels $K$ in Eq.~\eqref{eq:convolution} as well as the weights of the final dense layer. 
The input $F^{(0)}$ of the network 
is constructed by 
a configuration $(\si, \si')$. 
The network architecture is depicted in Figure~\ref{fig:architecture}, including a so-called pooling layer that is described below. 

Periodic boundary conditions (pbc) in the physical system can be considered by applying them to each convolutional layer's input. 
Even though over the length of the spin chain the same parameters are used repeatedly, 
the resulting feature maps for a given configuration are not translation invariant as there is still information about where in the lattice a feature occurred. 
Translation invariance for systems with periodic boundary conditions can be easily 
imposed 
using a single pooling layer on the output of the last convolution, averaging all nodes along the physical dimensions 
\begin{equation}
F^{(\text{pool})}_{k} = \frac{1}{XY} \sum_{x, y} F_{x,y,k} \, .
\label{eq:pooling}
\end{equation}
This greatly reduces the number of parameters and connections in the fully connected layer and, more importantly, the resulting network does not depend on the size of the input and hence the same parameter set can be used for any size of the physical system under consideration. This enables what is called \emph{transfer learning}, 
which amounts in  pre-training the model, for example on a smaller spin chain, 
 and using these parameters as the initial values for a larger system \cite{choo_study_2019}. 
In this 
way, the kernels can be trained to assume shapes that detect relevant features in the spin configurations and their correlations occurring in the steady-state, which are likely to also appear in larger physical systems. 
This often improves the obtained results and can accelerate convergence. 
In contrast to applying the translation operator to all input configurations and summing over the symmetry group members as is conventionally done \cite{nomura_helping_2021, nigro_invariant_2021}, with this architecture no additional effort is needed.


%

\section{Results} 

\subsection{Results for a dissipative transverse-field Ising model on a spin chain}

To evaluate the expressive power of the neural network ansatz described in Section.~\ref{sec:network}, 
we apply it to 
the problem of finding 
the non-equilibrium steady state (NESS) of the 1D dissipative transverse field Ising (TFI) 
model of $N$ spins with periodic boundary conditions. 
The Hamilton operator for this system is
\begin{equation}
H = \frac{V}{4}
\sum_{j} \sigma_j^z \sigma_{j+1}^z + \frac{g}{2} \sigma_j^x
\label{eq:tfising}
\end{equation}
with the Pauli matrices $\sigma_j^{x,y,z}$ at site $j$, an energy scale $V$ and a magnetic field strength $g$. 
The homogeneous dissipation is described 
$\gamma_k = \gamma$ (in Eq.~\eqref{eq:lindblad}) and by the jump operators $L_j = \sigma^-_j = \frac{1}{2}(\sigma_j^x - i \sigma_j^y)$ on all sites $j$. 
We set $V/\gamma = 2$ to compare with the results of Ref.~\cite{vicentini_variational_2019}.

\begin{figure}[h!]
	\centering
	\includegraphics[clip, trim=0cm 0cm 0cm 0cm,
	width=0.9\linewidth]
	{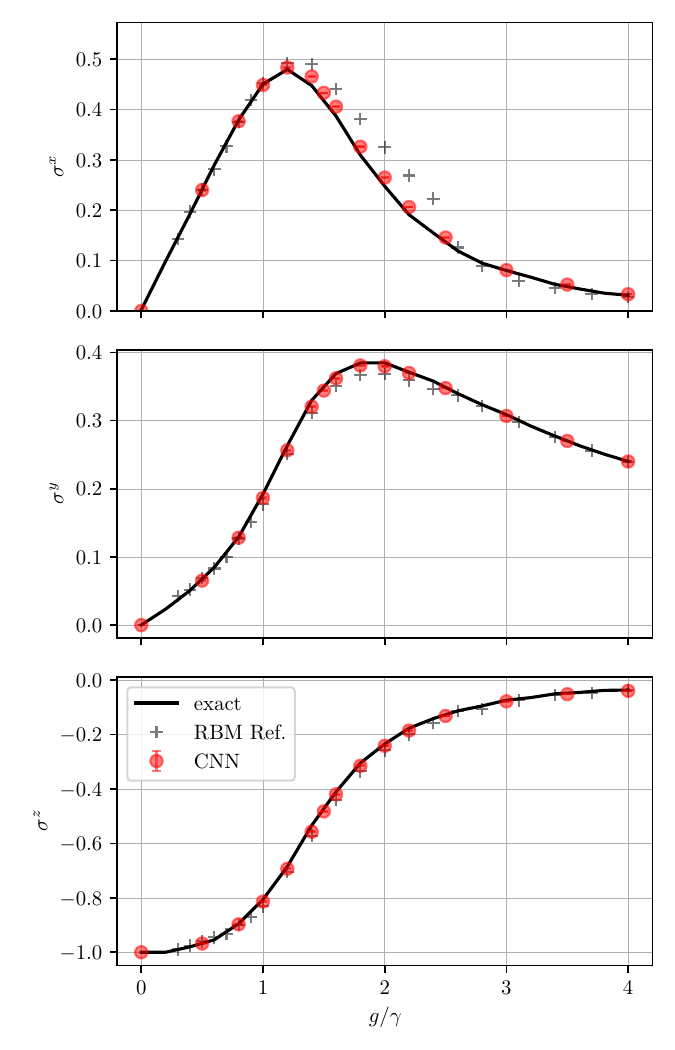} 
	\caption{One dimensional dissipative transverse field Ising model with $N=16$ and periodic boundary conditions. Results obtained by optimizing our CNN ansatz described in section~\ref{sec:network} 
compared to exact results and results obtained using RBM from Ref.~\cite{vicentini_variational_2019}. 
}
	\label{fig:results1d}
	\hfill
\end{figure}

For the spin chain we use the network architecture depicted in Fig.~\ref{fig:architecture}, using two convolutional layers with $6$ and $20$ feature maps with kernel sizes $(X,Y,C) = (3,1,2)$ and $(3,1,6)$ respectively, followed by a mean pooling over the spatial dimension and a fully connected final layer. 
In Fig.~\ref{fig:results1d} we plot the observables $\sigma^x$, $\sigma^y$ and $\sigma^z$ averaged over all sites 
for different magnetic field strengths $g$ using our CNN ansatz compared to exact values and the results obtained using RBM 
from Ref.~\cite{vicentini_variational_2019}. 
We can see that the CNN produces results with good accuracy, also in the range of the magnetic field $g / \gamma$ from $1$ to $2.5$, 
where the RBM had trouble producing the correct result even with an increased computational effort (compare Ref.~\cite{vicentini_variational_2019}).

%
This is achieved using only $438$ parameters, which is considerably less than the RBM which had $2752$ 
trainable parameters with hidden and ancillary node densities $1$ and $4$ respectively, as was used in Ref.~\cite{vicentini_variational_2019}. 
%
%
We chose a relatively small number of feature-maps in the first layer, as they tend to become redundant, whereas in consecutive layers, more parameters improve the results. 

We found that the initialization of the parameters of convolutional layers has a big impact on the performance. 
We initialize the kernels following a normal distribution with zero mean and a standard deviation $\sqrt{2 / v_n}$ with the number of parameters $v_n$ in the $n$-th layer, in order to control the variance throughout the network \cite{he_delving_2015}. 
We then further initialized the parameters by pre-training on a $6$-site chain. 
During optimization, the sums in Eq.~\eqref{eq:sum:HS} were evaluated using a sample size of $1024$ for $5000$ to $20000$ iterations until converged. The final observables were computed according to Eq.~\eqref{eq:observable} with $100000$ samples to reduce the variance. 

\begin{figure}[h!]
	\centering
	\includegraphics[clip, trim=0cm 0cm 0cm 0cm,
	width=0.9\linewidth]
	{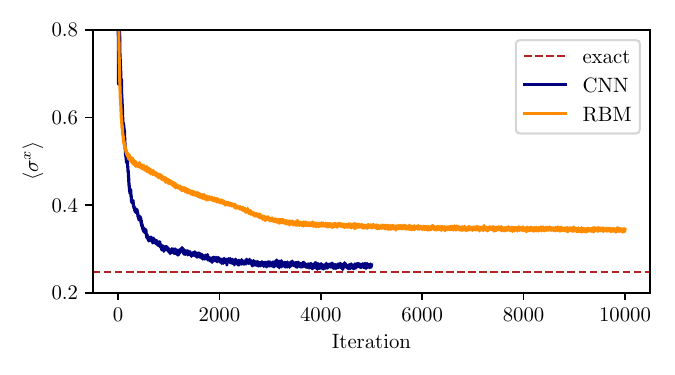} 
	\caption{Convergence of the $ \sigma^x $  observable for a $6$-spin dissipative TFI model with $g/\gamma = 2$ using a RBM as described in Ref.~\cite{vicentini_variational_2019} and the CNN architecture compared to exact diagonalization.}
	\label{fig:convergence_TFI_6}
	\hfill
\end{figure}

Interestingly, 
with RBM it was not possible (contrary to the CNN) to achieve an accurate result for certain parameter ranges with reasonable effort, even for such a small system of $6$ spins, as can be seen in Fig.~\ref{fig:convergence_TFI_6} where we compare $\langle \sigma^x \rangle$ during the optimization. 
This is probably due to the limited expressivity of the purified RBM description.

The constant number of parameters in principle makes it easier to scale to larger systems. 
For $30$ spins we obtained similar expectation values as depicted in Fig.~\ref{fig:architecture}, 
such as $\langle \sigma^x \rangle = 0.27$ at $g/\gamma = 2$. 
Since in this setup the exact solution of these observables has no strong dependence on the chain length, this result seems plausible. 
For the 1D spin chain we found comparable results with network architectures without the mean pooling layer, 
but then there are more parameters in the final dense layer depending on the spatial dimension of the input, which leads to a higher computational effort. 
The advantage of this architecture is that one 
can treat non-translation invariant systems, for example 
without periodic boundary conditions, 
such as an asymmetrically driven 
dissipative spin chain as in Ref.~\cite{mellak_quantum_2023}. 





\subsection{Results for the 2D dissipative Heisenberg model}

\begin{figure}[h!]
	\centering
	\includegraphics[clip, trim=0cm 0cm 0cm 0cm,
	width=0.9\linewidth]
	{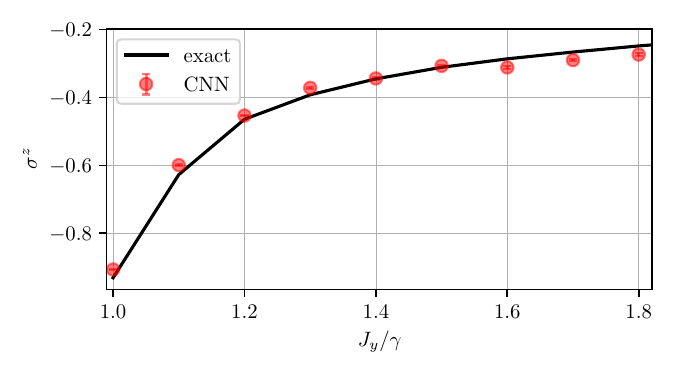} 
	\caption{Steady-state expectation values in a $3 \times 3$ dissipative Heisenberg model with periodic boundary conditions obtained by optimizing the CNN ansatz compared to exact values from Ref.~\cite{luo_autoregressive_2022}.} 
\label{fig:results2d}
\hfill
\end{figure}

To demonstrate expanding the network to higher dimensional systems, we look at the 2D dissipative Heisenberg model with periodic boundary conditions. 
The Hamiltonian reads 
\begin{equation}
	H = \sum_{\langle j, k \rangle} \left(J_x \; \sigma_j^x \sigma_{k}^x + J_y \; \sigma_j^y \sigma_{k}^y  + J_z \; \sigma_j^z \sigma_{k}^z \right) \; .
	\label{eq:Heisenberg}
\end{equation}
Following the setup in Ref.~\cite{luo_autoregressive_2022} a uniform dissipation rate $\gamma$ for the jump operators $L_j = \sigma^-_j$  and $J_x = 0.9 \gamma$, $J_z = \gamma$ are set.
In Fig.~\ref{fig:results2d} the steady-state results of the $\sigma^z$ expectation value for a $3\times3$ lattice is plotted for different values of $J_y / \gamma$, obtained by optimizing the CNN ansatz compared to exact values from Ref.~\cite{luo_autoregressive_2022}. 
Using only $350$ variational parameters, the CNN 
achieves comparable accuracy to the variational POVM solution by Ref.~\cite{luo_autoregressive_2022}. 
In their work  
they improved the variational results by running real-time evolution steps starting from the final optimized state. This could potentially present a method for further improving an already converged CNN result as well. 

\begin{figure}[h!]
	\centering
	\includegraphics[clip, trim=0cm 1cm 0cm 1cm,
	width=0.45\linewidth]
	{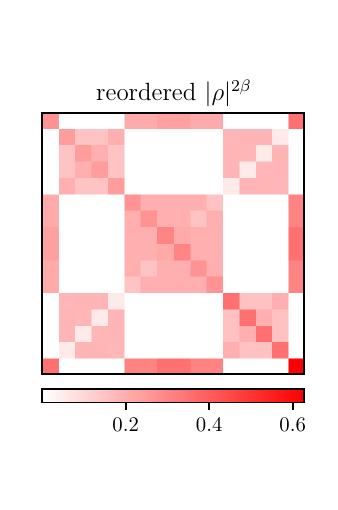} 
	\caption{
		Exact steady-state density matrix $\rho(\si, \si')$ of a $2 \times 2$ dissipative Heisenberg model, scaled as $|\rho|^{2\beta}$ with $\beta = 0.2$ to obtain a probability distribution that is easier to sample from and ordered according to the total spin of the configurations to show the allowed sectors.
	}
\label{fig:2dHeisRho0}
\hfill
\end{figure}

Here we chose a smaller $2 \times 2$ kernel size in the physical dimensions and $3$ convolutional layers with $6$ kernels each, followed by a mean pooling and a fully connected layer, analogues to Fig.~\ref{fig:architecture}. 
Due to the symmetry of the Hamiltonian and the Lindblad operators,
$\rho(\si, \si')$ is nonzero
only in sectors where $\sum_i (\sigma_i - \sigma'_i) = 2n$ with $n \in \mathbb{Z}$.
Following Ref.~\cite{mellak_quantum_2023}, this restriction can be implemented in the Monte-Carlo sampling by proposing only allowed configurations, leading to a faster convergence. We also rescaled the Monte-Carlo weights from Eq.~\ref{eq:prob_sig_sig'} using $|\rho|^{2 \beta}$ with $\beta = 0.2$ to better cover the configuration space, as described in Ref.~\cite{mellak_quantum_2023}. 
This new probability distribution is easier to sample from, as can be seen in Fig.~\ref{fig:2dHeisRho0}, where for a $2 \times 2$ lattice the rescaled exact density matrix $|\rho|^{2 \beta}$ is displayed, reordered according to the total spin of the configurations showing the allowed sectors. 
We again used a sample size of $1024$ during the optimization.


\section{Conclusion}

We demonstrated how a deep neural network ansatz can improve upon previous variational results 
by parametrizing the mixed density matrix directly and not enforcing the positivity. 
We rearranged the left and right Hilbert space of the spin configurations, which enabled a simple convolutional network architecture to efficiently capture the NESS of the dissipative transverse field Ising model with considerably less parameters compared to neural density operators ansatz functions based on RBM. 
These results encourage to explore other powerful neural network architectures to represent mixed density matrices without the 
explicit 
constraints of positivity, which RBM density operators were designed around.

Convolutions capture local correlations only, but varying the kernel sizes and the depth of the network, it can be tuned to better express longer range correlations. 
The simplicity of the network architecture makes it easily expandable and possibly interpretable, as the first-layer kernels for example should learn important spin-spin correlations that are then connected to each other in following layers. 

By introducing a pooling layer over all physical dimensions, translation invariance is enforced at no additional cost and the number of parameters is reduced at the same time. 
This, combined with the fact that in the convolutional layers there are no size-dependent fixed connections, enables transfer learning - using the same set of parameters as initialization for different physical system sizes. 
While this leads to a good fit for translation invariant models, it also encourages to look into CNN density matrices for physical models without such symmetries by leaving out the pooling step to keep the locality information in the network. 
Designing and applying complex valued networks and parameters could be another interesting area for further investigation.




\section*{Acknowledgments}
We would like to thank Thomas Pock for providing his expertise in insightful discussions. 
The implementation was based on the Jax \cite{jax2018github} and NetKet \cite{vicentini_netket_2022} libraries. 
This research was partly funded by the Austrian Science Fund (Grant No. P 33165-N) and by NaWi Graz. 

\newpage
\onecolumngrid

\twocolumngrid

\bibliographystyle{apsrev4-2}
\bibliography{NeuralQuantumNetwork,additionalReferences}

\end{document}